\documentclass[twocolumn, a4paper,10pt]{article}
\usepackage[top=2.5cm, bottom=2.5cm, left=2.0cm, right=2.0cm,
columnsep=0.8cm]{geometry}
\usepackage{enumitem}
\usepackage[hidelinks]{hyperref}
\usepackage{boxedminipage}
\usepackage{nopageno}
\usepackage{graphicx}
\usepackage{natbib}
\usepackage[font=it]{caption}
\usepackage[usenames,dvipsnames]{xcolor}
\usepackage{listings}
\usepackage{caption}
\usepackage{subcaption}

\usepackage{amsmath, amsfonts, amssymb, amstext, amsthm, dsfont}

\usepackage{nomencl}
\makenomenclature
\usepackage{etoolbox}
\renewcommand\nomgroup[1]{%
	\item[\bfseries
	\ifstrequal{#1}{L}{Latin letters}{%
		\ifstrequal{#1}{G}{Greek letters}{%
			\ifstrequal{#1}{D}{Parameters involved in the dimensionless representation}{%
				\ifstrequal{#1}{S}{Subscripts}{%
					\ifstrequal{#1}{U}{Superscripts}{}
				}%
			}%
		}%
	}%
	]}
\newcommand*\pd[2]{\frac{\partial #1}{\partial #2}}

\newcommand{\eqdef}{\mathop{\stackrel{\,\mathrm{def}}{:=}\,}}

\newcommand*\egal{\ = \ }
\newcommand*\plus{\ + \ }
\newcommand*\moins{\ - \ }

\newcommand{\BiT}{\mathrm{Bi}_{\,T}}
\newcommand{\BiM}{\mathrm{Bi}_{\,M}}

\newcommand{\cTs}{c_{\,T}}

\newcommand{\FoT}{\mathrm{Fo}_{\,T}}
\newcommand{\FoM}{\mathrm{Fo}_{\,M}}

\newcommand{\kTs}{k_{\,T}}
\newcommand{\kTMs}{k_{\,TM}}
\newcommand{\ts}{t}
\newcommand{\uinf}{u_{\,\infty}}
\newcommand{\uinfL}{u_{\,\infty}^{\,L}}
\newcommand{\uinfR}{u_{\,\infty}^{\,R}}
\newcommand{\vinf}{v_{\,\infty}}
\newcommand{\vinfL}{v_{\,\infty}^{\,L}}
\newcommand{\vinfR}{v_{\,\infty}^{\,R}}

\setlist{itemsep=-0.1cm,topsep=0.1cm,labelsep=0.3cm}
\setenumerate{leftmargin=*}
\makeatletter
\renewcommand\title[1]{\gdef\@title{\fontsize{12pt}{2pt}\bfseries{#1}}}
\makeatletter
\renewcommand\section{\@startsection{section}{1}{\z@}{3pt}{3pt}{\normalfont\large\bfseries}}
\makeatletter
\renewcommand\subsection{\@startsection{subsection}{1}{\z@}{\z@}{\z@}{\normalfont\normalsize\bfseries}}
\makeatletter
\renewcommand\subsection{\@startsection{subsection}{1}{\z@}{\z@}{0.1pt}{\normalfont\normalsize\bfseries}}

\usepackage{soul}
\definecolor{dkred}{rgb}{0.7,0,0}
\newcommand*\rev[1]{\textcolor{black}{#1}}

\title{%
	An efficient numerical method for a long-term simulation of heat and mass transfer: the case of an insulated rammed earth wall 
	\vspace{4pt}
} 																																		
\author{Madina Abdykarim$^1$, Julien Berger$^1$, Denys Dutykh$^2$, Amen Agbossou$^1$ \\ \\
	$^1$Univ. Grenoble Alpes, Univ. Savoie Mont Blanc, \\ 
	UMR 5271 CNRS, LOCIE, 73000 Chambery, France\\ 																																	
	$^2$Univ. Grenoble Alpes, Univ. Savoie Mont Blanc, \\ 
	UMR 5127 CNRS, LAMA, 73000 Chambery, France
}

\date{}

\begin{document}
	\maketitle
	\section*{Abstract}	
	\addtocounter{section}{1}
	Innovative numerical scheme studied in this work enables to overcome two main limitations of Building Performance Simulation (BPS) programs as high computational cost and the choice of a very fine numerical grid. 
	The method, called Super-Time-Stepping (STS), is novel to the state-of-the-art of building simulations, but has already proved to be sufficiently efficient in recent studies from anisotropic heat conduction in astrophysics (\cite{meyer2014}).
	The given research is focused on employment of this adopted numerical method to model drying of a rammed earth wall with an additional insulation layer. 	
	The results show considerable advantage of the STS method compared to standard \textsc{Euler} explicit scheme. 
	It is possible to choose at least $100$ times bigger time-steps to maintain high accuracy and to cut computational cost by more than $92 \, \%$ in the same time.
	\section*{Introduction}
	One of the main goals of any construction engineer is to avoid a possible damage. 
	Moisture is considered to be the most important source of natural destruction of building envelopes (\cite{guimaraes2018}).
	Various Building Performance Simulation (BPS) programs are used by practitioners in order to predict, simulate and analyze, among other phenomena, coupled heat and moisture transfer. 
	Nonetheless, state-of-the-art studies (\cite{clark2010ancient, hong2018}) highlight the need for innovative computational approaches, which may help to achieve a high computational accuracy with low costs while retaining the advantages of an explicit formulation.
	
	Despite its history of almost $40$ years (\cite{gentzsch1980}), the group of methods called Super--Time--Stepping (STS) \rev{requires attention for long-term simulations}.
	The STS allows to overcome two main limitations of traditional methods, namely the high computational cost and the choice of a very fine numerical grid. 
	The purpose of this article is the investigation of advantages of the STS method to perform long-term simulations of heat and moisture transfers through walls with an insulation layer on either sides of it. 
	
	Earth based materials are often considered to be a sustainable alternative. 
	They are also reusable and have low environmental impact (\cite{el2015numerical}). 
	According to their physical properties, these types of materials are a subject to drying and wetting during their lifetime. 
	So one can impose a question whether it is feasible to put an insulation layer together with a rammed earth (RE) wall or not. 
	In order to study the general impact of such configuration one needs to run a long-term simulation.  
	In this case, a faster numerical method comes in handy and, thereby, for this particular article, the implementation of the STS method will be extended for the multi-layered model and the strengths of the method will be investigated. 	
	
	The article is organized as follows. 
	The mathematical model of the physical phenomena is presented first. 
	The numerical method is described in the following section. 
	Verification of the theoretical results for the numerical schemes 
	as well as the numerical investigation with the real physical data are presented in the last two sections. 
	\section*{Mathematical Model}\label{sec:Physical_Model}
\begin{figure}
		\centering
		\includegraphics[width=0.5\textwidth]{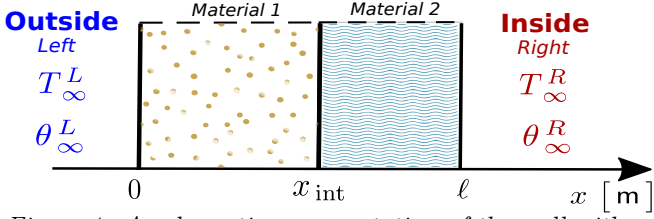}
		\caption{A schematic representation of the wall with an insulation layer.}
		\label{fig:BC_scheme}
	\end{figure}

	The section presents the mathematical model of one-dimensional heat and moisture transfer through porous material through the spatial $\Omega_{\, x} \egal [ \, 0, \, \ell \, ]$ and the time $\Omega_{\, t} \egal \left[ \, 0, \, \tau \, \right]$ domains, with $\ell \ \bigl[\,\mathsf{m}\,\bigr]$ being the total thickness of a wall and $\tau \ \bigl[\,\mathsf{h}\,\bigr]$ being the final time. 
	The wall is schematically illustrated in Figure~\ref{fig:BC_scheme}.
	The governing equations are based on energy and mass conservation equations (\cite{mendes2002}). 
	The subscripts $0\,$, $1$ and $2$ represent the dry state of the material, the water vapor and the liquid water, respectively. 
	The mass balance is written as follows:
	\begin{equation}	\label{eq:mass_balance}
	\rho_{\,2} \cdot \pd{ \theta}{ t} \egal - \, \pd{ j_{\,12}}{ x} \,,
	\end{equation}
	where $\rho_{\,2} \ \bigl[\,\mathsf{kg/m^{\,3}}\,\bigr]$ is the specific mass of liquid water and $\theta \ \bigl[\,\varnothing\,\bigr]$ is the volumetric moisture (liquid plus vapor) content. 
	The density of the moisture flow rate, $j_{\,12} \ \bigl[\,\mathsf{kg/(m^{\,2} \cdot s)}\,\bigr]$, includes the water vapor flow rate $j_{\, 1}$ and liquid water flow rate $j_{\, 2}$, so that $j_{\,12} \, \equiv \, j_{\, 1} \plus j_{\, 2}$.
	
	The internal heat conservation equation enables to state the temperature $T \ \bigl[\,\mathsf{K}\,\bigr]$ evolution law:
	\begin{equation}
	\label{eq:heat_balance}
	\Bigl(\, \rho_{\,0} \cdot c_{\,0} \plus \rho_{\,2} \cdot c_{\,2} \cdot \theta \,\Bigr) \cdot \pd{T}{t} \egal - \, \pd{j_{\,\mathrm{q}}}{x} \moins L_{\,12}^{\,\circ} \cdot \pd{j_{\,1}}{x} \,,
	\end{equation}
	where $\rho_{\,0} \ \bigl[\,\mathsf{kg/m^{\,3}}\,\bigr]$ is the specific mass of the dry material, $c_{\,0} \ \bigl[\,\mathsf{J /( kg \cdot K)}\,\bigr]$ is the material heat capacity and $c_{\,2} \ \bigl[\,\mathsf{J /( kg \cdot K)}\,\bigr]$ is the water heat capacity. 
	The quantity $j_{\,\mathrm{q}} \ \bigl[\,\mathsf{W/m^{\,2}}\,\bigr]$ is the sensible heat flow rate. 
	The latent heat of vaporization $L_{\,12}^{\,\circ} \ \bigl[\,\mathsf{J/kg}\,\bigr]$ is taken as a positive constant value.
	For the sake of clarity, we introduce the so-called global heat storage coefficient $\cTs \ \bigl[\,\mathsf{W\cdot s /( m^{\,3} \cdot K)}\,\bigr]\,$,  
	$ \cTs \, : \, \theta \, \mapsto \rho_{\,0} \cdot c_{\,0} \plus \rho_{\,2} \cdot c_{\,2} \cdot \theta \,.$
	Finally, the mathematical model can be expressed by the system of two coupled partial differential equations with respect to two unknowns $T$ and $\theta$:
	\begin{subequations}
		\label{eq:physical_model}
		\begin{align}
		\rho_{\,2} \cdot \pd{\theta}{t} & \, = \, \pd{}{x} \, \biggl(\, D_{\,\theta} \cdot \pd{\theta}{x} \, + \, D_{\,T} \cdot \pd{T}{x} \,\biggr) \,, \label{eq:mass_equation} \\[4pt]
		\cTs \cdot  \pd{T}{t} &\, = \, \pd{}{x} \, \biggl(\, k_{\,T} \cdot  \pd{T}{x} \, \biggr)  
		\, + \, L_{\,12}^{\,\circ} \cdot \pd{}{x} \, \biggl(\,  k_{\,TM} \cdot  \pd{\theta}{x} \, \biggr)  \,,
		\label{eq:heat_equation}
		\end{align}
	\end{subequations}
	where $D_{\,\theta} \, \left(\,T,\,\theta\,\right) \, \bigl[\,\mathsf{m^{\,2}/s}\,\bigr]$ is the diffusion coefficient under the moisture gradient, 
	$D_{\,T} \, \left(\,T,\,\theta\,\right) \, \bigl[\,\mathsf{m^{\,2}/(s \cdot K)}\,\bigr]$ is the diffusion coefficient under the temperature gradient, 
	$k_{\,T} \, \left(\,T,\,\theta\,\right) \, \bigl[\,\mathsf{W/(m \cdot K)}\,\bigr]$ is the thermal conductivity of the material, 
	and $k_{\,TM} \, \left(\,T,\,\theta\,\right) \, \bigl[\,\mathsf{kg/(m \cdot s)}\,\bigr]$ is the vapor transfer coefficient under the moisture gradient. 
	\subsection*{Multilayered domain}
	As the focus of this article is to study the influence of an insulation layer to the drying of a material, the natural continuity on the interface (\cite{freitas1996}) shall be applied. 
	Equation~\eqref{eq:physical_model} can be considered over a multidomain as illustrated in Figure~\ref{fig:BC_scheme}. 
	Both materials are taken as homogeneous and isotropic. 
	The space domain is written as $\Omega_{\, x} \egal  [ \, 0, \, x_{\, \text{int}} \,] \, \cup \, ] \, x_{\, \text{int}}, \,L \,]$ , 
	where $x_{\, \text{int}}$ is the location of the interface between two materials. 
	As a result, the material properties can be written in a general form as illustrated for the diffusion coefficient under
	moisture gradient:
	\begin{equation}
	D_{\,\theta} \, (\, \theta, \, T, \, x\,) \egal 
	\begin{cases}
	D_{\,\theta}^{\, \texttt{mat} 1} \, (\, \theta, \, T \,) \,, & \ x \, \leqslant \, x_{\, \text{int}} \,,\\
	D_{\,\theta}^{\, \texttt{mat} 2} \, (\, \theta, \, T \,) \,, & \ x \, > \, x_{\, \text{int}} \,, 
	\end{cases}
	\end{equation}
	where superscripts $\texttt{mat} 1$ and $\texttt{mat} 2$ represent each material layer. 
	
	One of the interesting quantities to study is the total moisture content remaining within the material, which can be  calculated as:
	\begin{equation}\label{eq:total_moisture_content}
	\theta^{\, \sf{tot}} \, (\,t \,)\eqdef \int_{\Omega_{\, \sf{Mat}}^{}} \ \theta \, (\, x,\, t \,)\ \mathrm{d} \, x \,,
	\end{equation}
	where $\Omega_{\, \sf{Mat}}$ is the domain of the material.  
	From this, one can also compute the rate of drying as the derivative of the total moisture content with respect to time:
	\begin{equation}\label{eq:velocity_drying}
	V^{\, \sf{dry}} \, (\,t \,) \eqdef \frac{\mathrm{d} \, \theta^{\, \sf{tot}} \, (\,t \,)}{\mathrm{d}\, t} \,.
	\end{equation}
	\subsection*{Boundary conditions}
	Assuming that there is no liquid water coming from the ambient environment, the boundary conditions at the surface $x \egal \left\{\,0, \,\ell \,\right\} $ for the moisture balance Equation~\eqref{eq:mass_balance} are written as follows:
	\begin{align}
	\label{eq:BC_mass_balance}
	&\left(\,D_{\,\theta} \cdot \pd{\theta}{{\bf n}} \plus D_{\,T} \cdot \pd{T}{{\bf n}} \,\right)  \egal   \\[4pt] \nonumber 
	& \frac{h_{\,M} \cdot M}{R_{\,1}} \cdot \Biggl(\, \varphi_{\, \infty} \cdot 
	\left(\, \frac{P_{\, \text{sat}}}{ \, T} \moins 
	\frac{P_{\,\text{sat},\, \infty}}{T_{\, \infty}} \,\right) \\[4pt] \nonumber 
	& \plus 
	\frac{P_{\, \text{sat}}}{ \, T} \cdot 
	\left(\,  \frac{d\, \widetilde{\varphi}}{d \,\theta} \cdot \Bigl(\, \theta \moins 
	\theta_{\, \infty} \,\Bigr) \plus r\,(\,\theta\,) \,\right) \,\Biggr) \, ,
	\end{align}
	where $h_{\,M}  \ \bigl[\,\mathsf{m/s}\,\bigr] $ is the surface vapor transfer coefficient, 
	$R_{\,1} \ \bigl[\,\mathsf{J/(\,kg \cdot\,K)}\,\bigr]$ is the constant gas for vapor, 
	$M \ \bigl[\,\mathsf{kg/mol}\,\bigr]$ is the molecular mass, 
	\rev{$\varphi \ \bigl[\,\varnothing\,\bigr]$ is the relative humidity},
	$P_{\, \text{sat}}\,(\,T\,) \eqdef 997.3 \cdot \, \left( \, \dfrac{T \moins 159.5 }{120.6}\, \right)^{\, 8.275} \ \bigl[\,\mathsf{Pa}\,\bigr]$ is the saturation pressure and
	$r \, \left(\,\theta \,\right)$ is the residual function as defined in \citet{mendes2002}. 
	$T_{\,\infty}$ and $\varphi_{\, \infty}$ stand for the temperature and the relative humidity of the ambient air.
	$\pd{g}{{\bf n}} \eqdef {\bf n} \cdot \pd{g}{x}$ is the directional derivative in the direction of the outer unit normal vector ${\bf n} \in \left\{\, -1 \,,\, 1 \,\right\}$, projected on the $O\, x$ axis. 
	
	The boundary conditions for the energy balance Equation~\eqref{eq:heat_balance} at the surface $x \egal \left\{\,0, \,\ell\,\right\} $:
	\begin{align}
	\label{eq:BC_heat_balance}
	&\left(\, k_{\,T} \cdot \pd{T}{{\bf n}} \, 
	\plus L_{\,12}^{\,\circ} \cdot k_{\,TM} \cdot \pd{\theta}{{\bf n}}  \,\right)
	\egal  \alpha \cdot g_{\, \infty} \\[4pt] \nonumber 
	& \, + \,  h_{\,T} \cdot \Bigl(\, T \, - \, T_{\,\infty} \,\Bigr) \, + \,
	L_{\,12}^{\,\circ}  \cdot \frac{h_{\,M}\cdot M}{R_{\,1}} \cdot \Biggl(\,
	\varphi_{\, \infty} \cdot \biggl(\, \frac{P_{\, \text{sat}}}{ \, T} \\[4pt] \nonumber 
	&\, - \,
	\frac{P_{\,\text{sat},\, \infty}}{T_{\, \infty}} \biggr) \, + \,
	\frac{P_{\, \text{sat}}}{ \, T} \cdot 
	\biggl(  \frac{d\, \widetilde{\varphi}}{d \,\theta} \cdot 
	\Bigl( \theta \, - \, \theta_{\, \infty} \,\Bigr) \, + \, r\,\left(\,\theta\,\right)\biggr) \Biggr)\,.
	\end{align}
	where $h_{\,T} \ \bigl[\,\mathsf{W/(m^{\,2} \cdot K)}\,\bigr]$ is the surface heat transfer coefficient and $\alpha \cdot g_{\, \infty} \ \bigl[\,\mathsf{W/m^{\,2}}\,\bigr]$ is the absorbed short-wave radiation. 

	\subsection*{Dimensionless formulation}
	The governing equations along with boundary conditions are solved numerically in a dimensionless form. 
	The solution in dimensionless formulation has advantages such as application to a class of problems sharing the same scaling parameters (e.g. \textsc{Fourier} and \textsc{Biot} numbers), simplification of a problem using asymptotic methods and restriction of round-off errors.  
	
	For the model given in Equation~\eqref{eq:physical_model}, the equations representing mass -- $v$ and heat -- $u$ transfer in porous material can be written in the dimensionless form for $x^{\, \star} \, \in \, \bigl[\,0 \,,\, 1\,\bigr]$ and $t^{\, \star} \, \in \, \bigl[\, 0 \,,\, \tau \,\bigr]$:
	\begin{multline}
	\pd{v}{t^{\, \star} }  \egal \FoM \cdot \pd{}{x^{\, \star} } \, \left(\, D_{\,\theta}^{\, \star} \cdot \pd{v}{x^{\, \star}} \plus \gamma \cdot  
			D_{\,T}^{\, \star} \cdot \pd{u}{x^{\, \star}} \,\right)\,, \\
				\cTs^{\, \star} \cdot \pd{u}{t^{\, \star}}  \, = \,
			\FoT \cdot \Biggl( \pd{}{x^{\, \star} }  \biggl( \kTs^{\, \star}\cdot \pd{u}{x^{\, \star}}\biggr) 
			\\
			\,+ \, 
			\delta \cdot \pd{}{x^{\, \star} }  \biggl( \kTMs^{\, \star} \cdot \pd{v}{x^{\, \star}} \biggr) \Biggr),
			\label{eq:dimless_model}
	\end{multline}
	where the superscript $^{\,\star}$ represents a dimensionless value of a variable \rev{ and $\gamma, \delta$ are dimensionless coupling parameters.}
	
	The initial conditions at $t^{\, \star} \, = \, 0 $ are $u_{\, 0} \, = \, v_{\, 0} \, = \, 1$ for $\forall \, x^{\, \star} \, \in \, \bigl[\,0 \,,\, 1\,\bigr]$.
	The boundary conditions at the surface $x^{\, \star} \egal \left\{\,0, \,1\,\right\}$ are defined as: 
	\begin{subequations}
		\label{eq:BC_dimless_model}
		\begin{align}
		&\left(\, D_{\,\theta}^{\, \star} \cdot \pd{v}{{\bf n}^{\, \star}} \plus \gamma \cdot D_{\,T}^{\, \star} \cdot \pd{u}{{\bf n}^{\, \star}} \,\right)
		\egal  G_{\, M}  \plus \\[4pt] 
		& \BiM^{\,\text{sat}} \cdot  \Biggl(\, \frac{P_{\, \text{sat}}^{\, \star}}{u} \moins \frac{P_{\,\text{sat} \,,\, \infty}^{\, \star}}{\uinf} \,\Biggr) \,  \nonumber  \plus 
		\BiM^{\,\theta} \cdot \Bigl(\, v \moins \vinf \,\Bigr)\,,
		\label{eq:BC_dimless_mass}  \nonumber \\[4pt]
		&\left(\, \kTs^{\, \star} \cdot \pd{v}{{\bf n}^{\, \star}} \plus \delta \cdot \kTMs^{\, \star} \cdot \pd{u}{{\bf n}^{\, \star}} \,\right)
		\egal  G_{\, T} \plus \\[4pt] 
		&\BiT^{\,T}\cdot \Bigl(\, u \moins \uinf \,\Bigr)
		\plus 
		\BiT^{\,\text{sat}} \cdot \Biggl(\, \frac{P_{\, \text{sat}}^{\, \star}}{u} \moins \frac{P_{\,\text{sat} \,,\, \infty}^{\, \star}}{\uinf} \,\Biggr)  \nonumber \\[4pt] 
		&  \plus 
		\BiT^{\,\theta} \cdot  \Bigl(\, v \moins \vinf \,\Bigr) \plus \alpha \cdot \BiT^{\,g} \cdot g_{\, \infty}^{\, \star} \,,  \nonumber
		\label{eq:BC_dimless_heat} 
		\end{align}
	\end{subequations}
	\rev{where $G_{\, M}, G_{\, T} $ are dimensionless additional thermal flux terms of the boundary conditions.}
	
	\rev{The \textsc{Fourier} numbers $\FoM$ and $\FoT$ are defined as}:
	\begin{equation*}
	\FoM  \ \eqdef \ \frac{t^{\,\circ} \cdot \, D_{\,\theta}^{\,\circ}}{\ell^2 \cdot \, \rho_{\,2}}\,, 
	\qquad
	\FoT \ \eqdef \ \frac{t^{\,\circ} \cdot \, \kTs^{\,\circ}}{\ell^2 \cdot \, \cTs^{\,\circ}}\,.
	\end{equation*}
	\rev{The \textsc{Biot} numbers $\BiM^{\,\text{sat}}, \BiM^{\,\theta}, \BiT^{\,T}, \BiT^{\,\text{sat}}, \BiT^{\,\theta}$ and $\BiT^{\,g}$ can be expressed as}:
	\begin{equation*}
	\begin{split}
	\BiM^{\,\text{sat}} &\ \eqdef \  \frac{\ell \, \cdot \, \varphi_{\, \infty}^{\,\circ} \, \cdot \, h_{\,M} \cdot M}
	{D_{\,\theta}^{\,\circ} \cdot \, \theta^{\,\circ}  \, \cdot \, R_{\,1}} 
	\, \cdot \, \frac{P_{\, \text{sat}}^{\,\circ}}{T^{\,\circ}} \cdot \varphi_{\, \infty}^{\, \star}\,, \  \\[4pt]
	\BiM^{\,\theta} &\ \eqdef \  \frac{\ell \, \cdot \, h_{\,M}\cdot M}{D_{\,\theta}^{\,\circ} \cdot \, R_{\,1}}  
	\, \cdot \, \frac{P_{\, \text{sat}}^{\,\circ}}{T^{\,\circ}} 
	\, \cdot \, \frac{d\, \widetilde{\varphi}}{d \,\theta} 
	\, \cdot \, \frac{P_{\, \text{sat}}^{\, \star}}{u} \,,\ \\[4pt]
	\BiT^{\,T} &\ \eqdef \  \frac{\ell \, \cdot \, h_{\,T} }{\kTs^{\,\circ}} \,,\qquad
	\BiT^{\,g} \ \eqdef \  \frac{\ell \, \cdot \, g_{\, \infty}^{\,\circ} }{\kTs^{\,\circ} \cdot T^{\,\circ}} \,, \\[4pt]
	\BiT^{\,\text{sat}} &\ \eqdef \  L_{\,12}^{\,\circ} \cdot \,  \frac{\ell \, \cdot \, \varphi_{\, \infty}^{\,\circ} \, \cdot \, h_{\,M} \cdot M}
	{\kTs^{\,\circ} \cdot \, T^{\,\circ}  \, \cdot \, R_{\,1}} 
	\, \cdot \, \frac{P_{\, \text{sat}}^{\,\circ}}{T^{\,\circ}} \; \cdot \varphi_{\, \infty}^{\, \star}\,,\  \\[4pt]
	\BiT^{\,\theta} &\ \eqdef \  L_{\,12}^{\,\circ} \cdot \,
	\frac{\ell \, \cdot \, \theta^{\,\circ} \, \cdot \, h_{\,M} \cdot M}{\kTs^{\,\circ} \cdot \, T^{\,\circ}\cdot \, R_{\,1}}  
	\, \cdot \,\frac{P_{\, \text{sat}}^{\,\circ}}{T^{\,\circ}} 
	\, \cdot \, \frac{d\, \widetilde{\varphi}}{d \,\theta} 
	\, \cdot \, \frac{P_{\, \text{sat}}^{\, \star}}{u}\,. 
	\end{split}
	\end{equation*}
	The next section presents the description of the STS method, which is proposed to be applied to the heat and moisture transfer simulation. 
	\section*{Numerical Methods}\label{sec:Numerical_Methods}
	For the sake of simplicity and without losing generality, in order to explain numerical schemes, the initial-boundary value problem is considered:
	\begin{equation}\label{eq:diffusion_equation}
	\pd{\, u}{\, t} \egal \pd{}{\, x} \, \Biggl( \, d \cdot \pd{\, u}{\, x}  \, \Biggr) \,,
	\end{equation}
	where $d$ is the material diffusivity. The initial condition is $u \bigl( x,\,t = 0 \bigr) \eqdef u_{\, 0} \, (\, x\,)$ and the boundary conditions are $u\, \bigl(\, x = \left\{\,0, \,1\,\right\}, \, t \, \bigr) \eqdef u_{\, \infty}^{\,L, \, R} \, (\, t\,) $. 
	
	The space and time domain are discretized in the following way. 
	A uniform discretization of the space interval $\Omega_{\, x}  \rightsquigarrow \Omega_{\, h} $ is written as $\Omega_{\, h} \egal \bigcup_{\, j \egal 1}^{\, N_{\, x}} [\, x_{\, j}, \, x_{\, j \plus 1}\, ], \quad x_{\, j \plus 1} \moins x_{\, j} \ \equiv \ \Delta \, x, \, \, \forall \, j \, \in \, \{ \, 1, \ldots, N_{\, x}\, \} \,. $
	Time layers are spaced uniformly as well $t^{\, n} \egal n \, \Delta \, t, \, \quad \Delta \, t = \textnormal{const} \ > \ 0, \, \, \forall \, n \, \in \, \{ \, 0, \ldots, N_{\, t} \, \}\,. $
	The values of the solution function $u\,(\, x,\, t\, )$ are defined at discrete nodes and denoted by $u_{\,j}^{\,n} \ := \ u\, (\, x_{\,j},\, t^{\,n} \,)$. 
	\subsection*{The Super--Time--Stepping Method}\label{sec:STS_Methods}
	Almost $40$ years ago the Super--Time--Stepping (STS) numerical method was proposed to solve parabolic problems by \cite{gentzsch1980}. 
	Since then, it was only employed for a limited range of problems and notable applications to linear and nonlinear parabolic problems were performed in \cite{alexiades1996}. 
	Those applications once again confirmed obvious advantages of the STS method as a tool to speed up remarkably the explicit time-stepping schemes in a very simple way. 
	The philosophy of the method lies in its \textsc{Runge--Kutta}-like nature.  
	The iterative algorithms of such methods, based on the recursion relations of orthogonal polynomials, permit ensuring the stability of the method at the end of each iteration stage.
	In this way, the numerical scheme is able to relax the strong stability requirement at the end of every small time-step. 
	The stability is then required only at the end of a cycle of $N_{\, \mathrm{S}}$ of them, where $N_{\, \mathrm{S}}$ is the number of super-time-steps.  
	\begin{figure}
		\centering
		\includegraphics[width=0.5\textwidth]{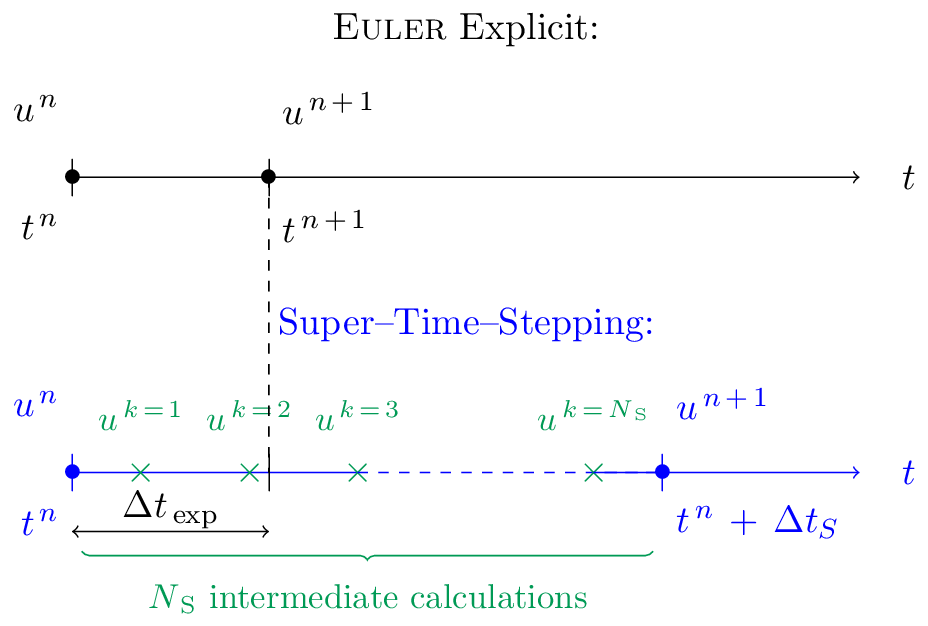}
		\caption{\small Stencil of the Super--Time--Stepping method in comparison with \textsc{Euler} explicit scheme (\ref{fig:STS_Stencil}).}
		\label{fig:STS_Stencil}
	\end{figure}
	In this article, two STS approaches are considered based on two families of orthogonal polynomials. 
	Namely, shifted \textsc{Chebyshev} polynomial of degree $N_{\,\mathrm{S}}$ (\cite{alexiades1996}) and shifted \textsc{Legendre} polynomials of the first order (\cite{meyer2014}).
	The general idea of the method is described below and additional details may be found in \cite{abdykarim2019}. 
	
	The \textsc{Euler} explicit discretization for the time-dependent linear diffusion Equation~\eqref{eq:diffusion_equation} can be written as:
	\begin{equation}\label{eq:STS_simple_heat_discretisation}
	u^{\, n \plus 1} \egal \bigl(\, \mathds{I} \moins \Delta \, t \cdot \mathds{A} \, \bigr) \cdot u^{\, n} \, , \quad n \, \in \, \mathds{N} \, ,
	\end{equation}
	where matrix $\mathds{A}$ can be constructed according to the chosen space discretization. 
	The stability condition of the scheme is associated to the spectral radius $\rho$ of the matrix operator:
	\begin{equation} \label{eq:stability_STS}
	\rho \, \bigl(\, \mathds{I} \moins \Delta \, t \cdot \mathds{A} \, \bigr) \ < \ 1 \, ,
	\end{equation}
	where the spectral radius operator $\rho\,(-)$, which is defined as:  
	\begin{align*}
	\rho \,: \ \mathrm{Mat}_{\,m \times m}\,(\mathds{R})\ &\longrightarrow\ \mathds{R}_{\,\geqslant\, 0}\,, \\[4pt]
	A\ &\longmapsto\ \max_{1\leqslant j\leqslant m}\Bigl\{\,\vert\,\lambda_{\,j}\,\vert\ \bigl\vert\ A\,v_{\,j}\ =\ \lambda_{\,j}\,v_{\,j}\,, \\[4pt] & \, v_{\,j}\ \in\ \mathds{R}^{\,m}\setminus{\boldsymbol{0}}\Bigr\}\ \in\ \mathds{R}_{\,\geqslant\, 0} \,.
	\end{align*}
	Note the maximum $\lambda_{\, \max}\,$ and the smallest $\lambda_{\, \min} \ > \ 0 $  eigenvalues of the matrix $\mathds{A}$. 
	The above relationship yields to the following stability condition of \textsc{Courant--Friedrichs--Lewy} (CFL) type for the time discretization:
	\begin{equation}\label{eq:STS_stability}
	|\, 1 \moins \Delta \, t \cdot \lambda_{\, \max}\, | \ < \ 1 \quad \implies \quad
	\Delta t \ < \ \Delta t_{\, \text{exp}}\,, 
	\end{equation}
	with
	\begin{equation}\label{eq:STS_delta_t_exp}
	\Delta t_{\, \text{exp}} \eqdef \frac{2}{\lambda_{\, \max}}\, ,
	\end{equation}
	and $\lambda_{\,\max} \egal  \dfrac{4 \,k }{\Delta \,x^{\, 2}}$ (\cite{alexiades1996}).
	
	The above condition~\eqref{eq:STS_stability} can be relaxed by introducing a stability polynomial $\mathsf{P}_{\, N_{\,\mathrm{S}}}$, $\forall \, \lambda \ \in \ [\, \lambda_{\, \min}, \, \lambda_{\, \max} \,]$:
	\begin{equation}\label{eq:STS_stability_polynomial}
	\Big | \, \mathsf{P}_{\, N_{\,\mathrm{S}}} \bigl(\, \Delta \, t_{\, S} \, , \lambda \, \bigr) \, \Big | \ \leqslant \ 1 \, .
	\end{equation}
	Here, it is possible to relax the stability constraint on each time-step $\Delta \, t$ by introducing a so-called super-time-step $\Delta \, t_{\, \mathrm{S}}$. 
	The stability is then required only at the end of a cycle of $N_{\, \mathrm{S}}$ super-time-steps. 
	This leads to the numerical scheme similar to a \textsc{Runge--Kutta}-like method with $N_{\, \mathrm{S}}$ stages. 
	The stencil of the STS scheme is shown in Figure~\ref{fig:STS_Stencil} to understand the technique idea. 
	One can observe that the STS method performs sequences of $N_{\, \mathrm{S}}$ inner steps (intermediate calculations) and in total performs $N_{\, \mathrm{S}} \cdot \dfrac{\tau}{\Delta\, t_{\, \mathrm{S}}}$ explicit steps, where $\tau$ is the final simulation time.  
	As a result, approximately $N_{\, \text{STS}} \eqdef \dfrac{\tau}{\Delta\, t_{\, \mathrm{S}}}$ temporal nodes are obtained.
	
	Thereby, one can express discretization~\eqref{eq:STS_simple_heat_discretisation} in the following way for $n \egal 0\,, 1\,, \ldots\, N_{\, STS}$:
	\begin{equation}\label{eq:STS_general_heat_equation_discretisation}
	u^{\, n+1} \egal \Biggl( \, \mathsf{P}_{\, N} \bigl(\, \Delta \, t_{\,\mathrm{S}} \,, \mathds{A} \, \bigr) \, \Biggr) \cdot u^{\, n} \,.
	\end{equation}
	Now the solution can be found with the scheme involving a super-time-step $\Delta \, t_{\,\mathrm{S}}$, which should satisfy either the \textsc{Chebyshev} or \textsc{Legendre} stability polynomials.
	Depending on the choice of a method, $\Delta \, t_{\,\mathrm{S}}$ can be fixed according to the number of super-time-steps $N_{\, \mathrm{S}}$ and explicit time-step $\Delta \, t_{\, \text{exp}}$ defined in~\eqref{eq:STS_delta_t_exp}: 
	
	\begin{itemize}
		\item \textbf{RKC}: \textsc{Runge--Kutta--Chebyshev} STS method: 
		\begin{equation}
		\Delta \, t_{\, \mathrm{S}} \egal \displaystyle\sum_{\, k \egal 1}^{\, N_{\, \mathrm{S}}} \tau_{\, k} \ \xrightarrow{\lambda_{\, \text{max}} }  \ N_{\, \mathrm{S}}^{\, 2}\cdot \Delta \, t_{\, \text{exp}}\,,
		\label{eq:STS_RKC_delta_tS}
		\end{equation}
		where $\tau_{\, k}$ is the time-step of intermediary stage $k$. 
		\item \textbf{RKL}: \textsc{Runge--Kutta--Legendre} STS method of the first order:
		\begin{equation}
		\Delta \, t_{\, \mathrm{S}} \ \leqslant \ \frac{N_{\, \mathrm{S}}^{\, 2} \plus N_{\, \mathrm{S}}}{2} \cdot \Delta \, t_{\, \text{exp}}\,. 
		\label{eq:STS_RKL_delta_tS}
		\end{equation}
	\end{itemize}
	
	As it can be seen, the super-time-step $\Delta \, t_{\,\mathrm{S}}$ can be at least $\mathcal{O} \, (\,N_{\, \mathrm{S}}^{\, 2}\,)$ bigger than a time-step required by the explicit \textsc{Euler} scheme due to the CFL stability condition~\eqref{eq:STS_stability}. 
	The expectations of a much faster calculation are based on this fact of a time-step ``widening". 
	A few case studies below will prove such effectiveness of the STS method in a variety of ways, which are described in the next section. 
	\subsection*{Comparing numerical results}
	In order to compare the efficiency of the method, results have been compared with the explicit \textsc{Euler} scheme and with the improved explicit method, called \textsc{{Du\,Fort}--Frankel} (DF) (for more information about DF method, readers can refer to \cite{du1953stability, gasparin2017stable}. 	
	
	Numerical methods can be compared by computing the $\varepsilon_{\,2}$ error between a numerical solution $u_{\, \text{num}}$ and the reference solution $u_{\, \text{ref}} \, $. 
	The accuracy can be estimated with the global uniform error \rev{$\varepsilon_{\,\infty}$} and the significant correct digits (scd) of a solution (\cite{gasparin2017stable, abdykarim2019}).

	To evaluate the efficiency of the methods in comparison with the explicit \textsc{Euler} scheme, the ratio of the total number of temporal steps $\varrho_{\, N_{\, \Delta \, \ts}} \ \bigl[\,\%\,\bigr]$ and the ratio of computational cost $\varrho_{\, \text{\tiny CPU}} \ \bigl[\,\%\,\bigr]$ can be computed as follows: 
	\begin{equation*}\label{eq:ntsr}
	\varrho_{\, N_{\, \Delta \, \ts}} \ \eqdef \dfrac{N_{\, \Delta \, \ts}^{\, \text{\tiny scheme}}}{N_{\, \Delta \, \ts}^{\, \textsc{\tiny Euler}}} \cdot 100 \% \,, \quad
	\varrho_{\, \text{\tiny CPU}} \ \eqdef \dfrac{t_{\, \text{\tiny CPU}}^{\, \text{\tiny scheme}}}{t_{\, \text{\tiny CPU}}^{\, \textsc{\tiny Euler}}} \cdot 100 \% \,,
	\end{equation*}
	where $N_{\, \ts}^{\, \text{\tiny scheme}}, \, t_{\, \text{\tiny CPU}}^{\, \text{\tiny scheme}} \ \bigl[\,\sf s\,\bigr]$ and $N_{\, \ts}^{\, \textsc{\tiny Euler}}, \,t_{\, \text{\tiny CPU}}^{\, \textsc{\tiny Euler}} \ \bigl[\,\sf s\,\bigr]$ are the total numbers of temporal steps and computational times required by the DF or STS schemes and by the \textsc{Euler} explicit scheme respectively.
	
	One can also calculate the computational time ratio per day $\varrho_{\, \text{\tiny CPU}}^{\, \text{\tiny day}} \ \eqdef \dfrac{t_{\, \text{\tiny CPU}}^{\, \text{\tiny scheme}}}{\tau_{\, \sf d}} \ \bigl[\,\sf s/d\,\bigr]$, which evaluates how many seconds are required to perform the simulation for one astronomical day. 
	
	In the following section, the numerical methods shall be validated with the reference solution and compared among each other. 
	\section*{Numerical verification}\label{sec:Numerical_Validation}
	For the first case, material properties are considered to be constant throughout materials and independent of the field of temperature and relative humidity.
	It is required to verify the theoretical results of the numerical scheme. 
	The model is taken in its dimensionless form as~\eqref{eq:dimless_model} together with the initial and boundary conditions \eqref{eq:BC_dimless_model}.
	
	Material properties are given in Table~\ref{tab:MatProp_dimless}  and also expressed with \textsc{Fourier} numbers, $\gamma$ and $\delta$ which are equal to $\FoT  \egal 7 \, \cdot 10^{\, -2},$ $ \FoM  \egal 9 \, \cdot 10^{\, -2},$ $\gamma \egal 7 \, \cdot 10^{\, -2}$ and $\delta \egal 5\cdot 10^{\, -2}$.  
		\begin{table}
		\centering  
		\small    
		\caption{\small Dimetionless material properties of two materials.}                                            
		\bigskip \label{tab:MatProp_dimless}  
		\begin{tabular}{|l|c|c|c|c|c|} 
		\cline{2-6}                                                                                
		\multicolumn{1}{c|}{}    &  $D_{\, \theta}^{\, \star} $ & $D_{\, T}^{\, \star} $ & $\cTs^{\, \star} $ & $\kTs^{\, \star} $ & $\kTMs^{\, \star} $\\                                              
		\hline \hline                                                                       
		\texttt{mat} 1 & $0.3$ & $2.1$& $0.1$& $0.5$& $0.4$ \\
		\hline \hline                                                                       
		\texttt{mat} 2 & $0.1$  & $3.2$ & $0.3$ & $0.2$ & $0.1$ \\
		\hline \hline                                                                     
		\end{tabular}                                                
		\end{table} 
		
	\textsc{Biot} numbers are expressed as parameters for the boundary conditions and are taken to be equal to:
	\begin{align*}
	\BiM^{\,\theta \,, L} &\, = \, 25.5\,, \ \ \BiT^{\,T\,, L} \, = \,  50.5\,, \ \  \BiT^{\,\theta\,, L}  \, = \,  4.96\cdot 10^{\, -1}\,, \\[4pt]
	\BiM^{\,\theta\,, R} &\, = \,  51.8\,, \ \ \BiT^{\,T\,, R} \, = \,  19.8\,, \ \ \BiT^{\,\theta\,, R}  \, = \,  6.73\cdot 10^{\, -1}\,,
	\end{align*}
	and all $\BiM^{\,\text{sat}} \, = \, \BiT^{\,\text{sat}} = 0$. 
	Additional flux parameters and the short-wave radiation are also set to zero. 
	The initial conditions for $u$ and $v$ are identically equal to one. 
	Variation of the boundary data is set to obey te following periodic functions:
	\begin{align*}
	\uinfL &\, = \,  1 \, + \, \frac{3}{5} \, \sin \left( 2\pi \frac{\ts}{5}\right)^{\,2} , \ \  \vinfL \, = \,  1 \, + \, \frac{1}{5} \, \sin \left( 2\pi \frac{\ts}{2}\right)^{\,2},\\[4pt]
	\uinfR &\, = \, 1 \, + \, \frac{1}{2} \, \sin \left( 2\pi \frac{\ts}{3}\right)^{\,2} , \ \  \vinfR \, = \,  1 \, + \, \frac{9}{10} \, \sin \left( 2\pi \frac{\ts}{6}\right)^{\,2}. 
	\end{align*}
		
	The total simulation time is $\tau^{\, \star}  \egal 1$. 
	The space discretization parameter is $\Delta \, x^{\, \star} \egal 10^{\,-2}$ for all schemes. 
	Interface between materials is placed to be at $x_{\, \text{int}}^{\, \star} \egal 0.6$.  
	The value of the time-step parameter is chosen according to corresponding requirements of each scheme. 
	The number of super-time-steps have been taken as $N^{\, \text{RKC}}_{\, \mathrm{S}} \egal 10$ and $N^{\, \text{RKL}}_{\, \mathrm{S}} \egal 20$.
	\subsection*{Results and discussion}
\begin{figure*}
		\centering
		\begin{subfigure}{0.47\textwidth}
			\includegraphics[width=1\textwidth]{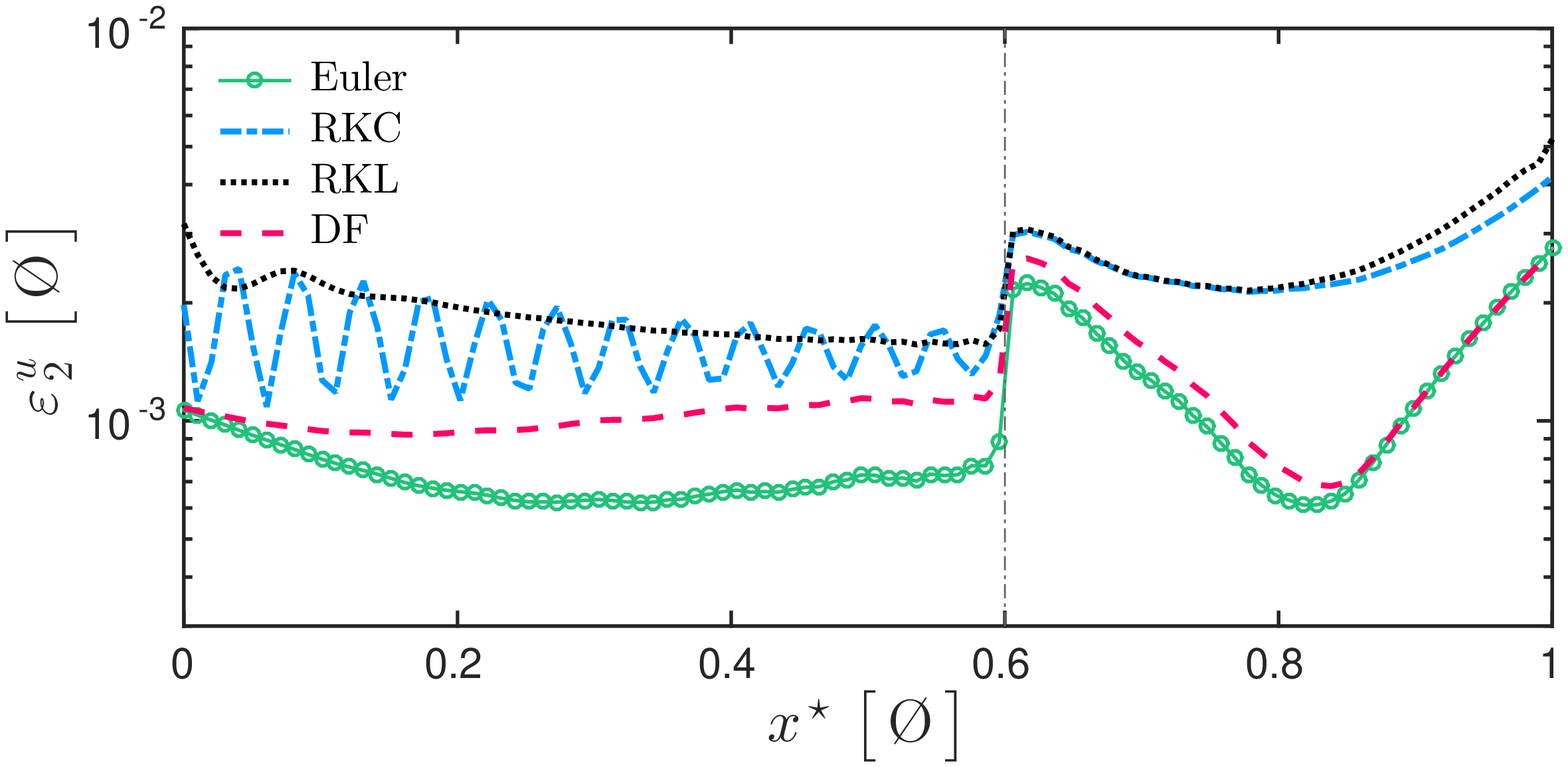} %
		\end{subfigure}
		\begin{subfigure}{0.47\textwidth}
			\includegraphics[width=1\textwidth]{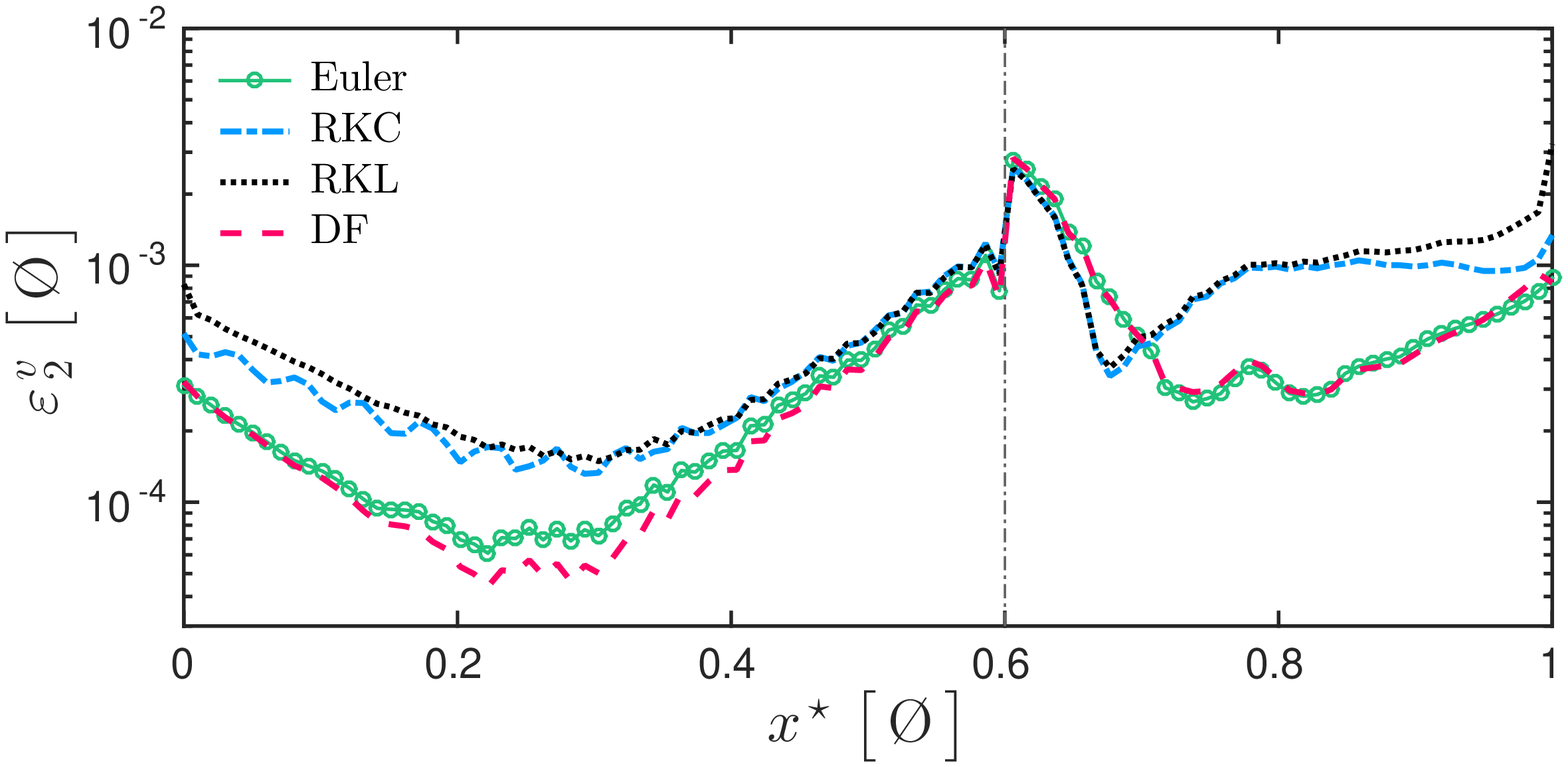} %
		\end{subfigure}
		\caption{\small  Error between the numerical simulation results and the reference solution, for respective time steps $\Delta \, t^{\,\star}$ reported in Table~\ref{tab:case_study} and $\Delta \, x^{\,\star} \egal 10^{\, -2}$. }
		\label{fig:Linear_case_study_error}
	\end{figure*}

\begin{figure*}
		\centering
		\begin{subfigure}{0.47\textwidth}
			\centering
			\includegraphics[width=1\textwidth]{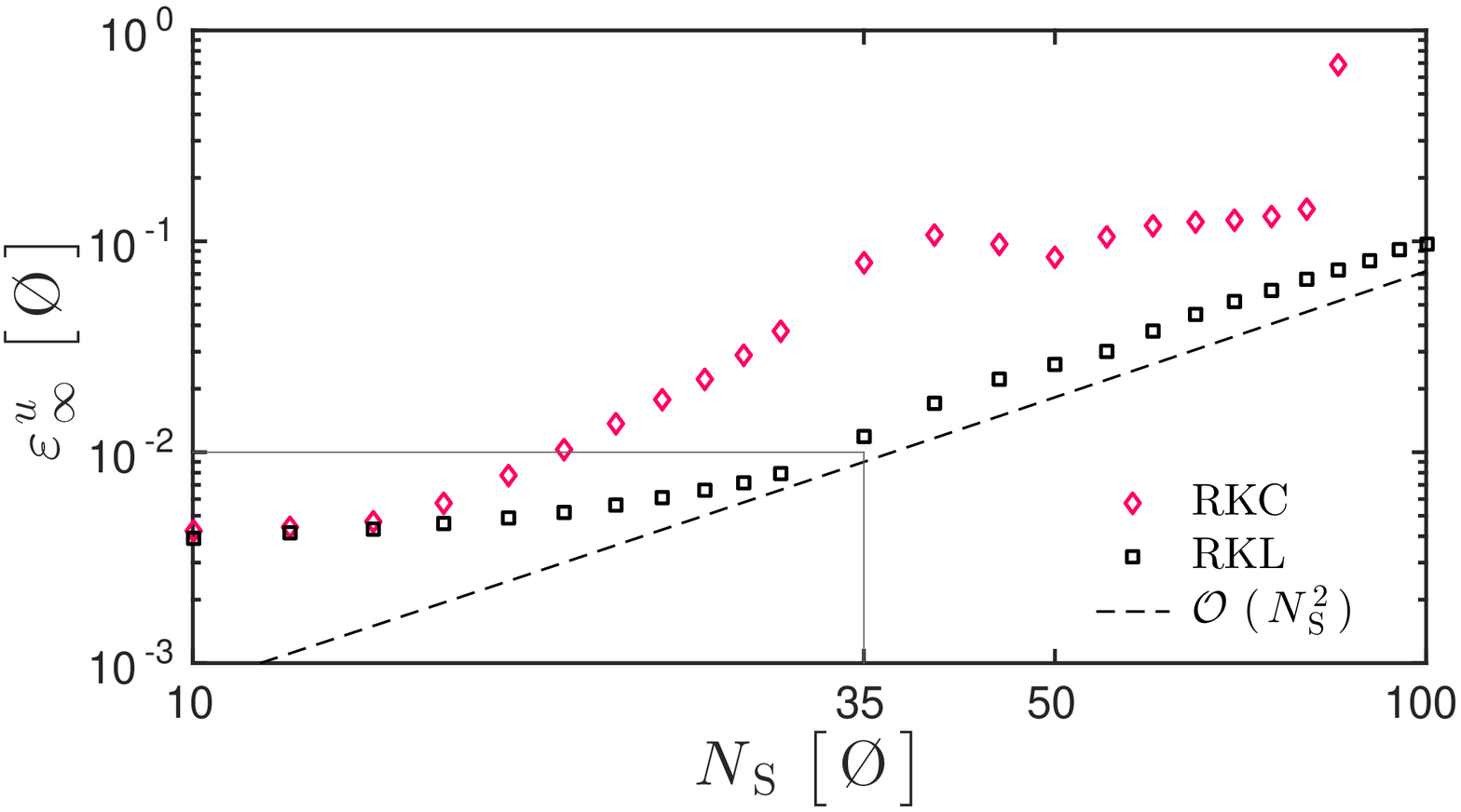}
			\caption{}
			\label{fig:Case_Study_Linear_errorSTS_U}
		\end{subfigure}
		\begin{subfigure}{0.47\textwidth}
			\centering
			\includegraphics[width=1\textwidth]{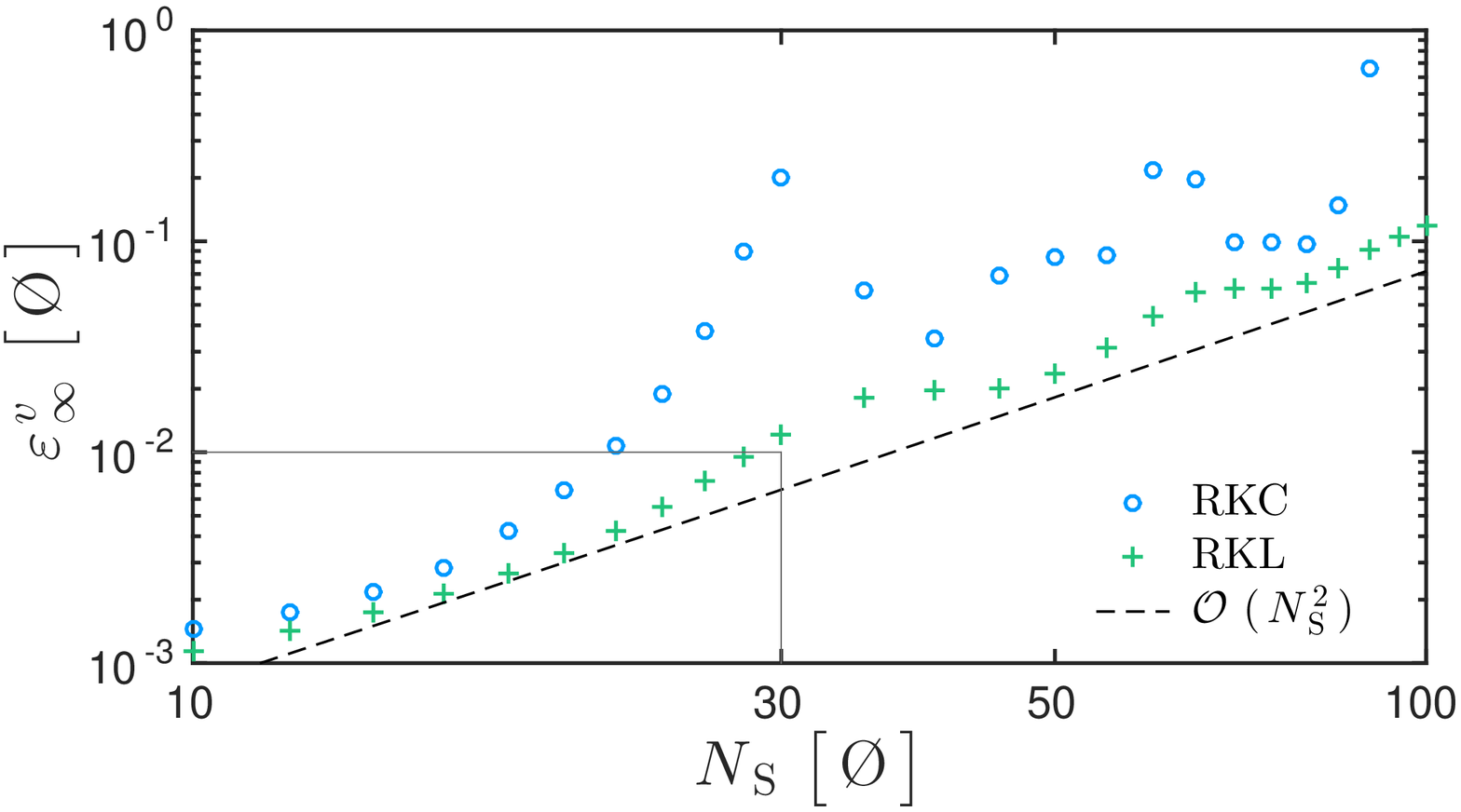}
			\caption{}
			\label{fig:Case_Study_Linear_errorSTS_V}
		\end{subfigure}
		\begin{subfigure}{0.47\textwidth}
			\centering
			\includegraphics[width=1\textwidth]{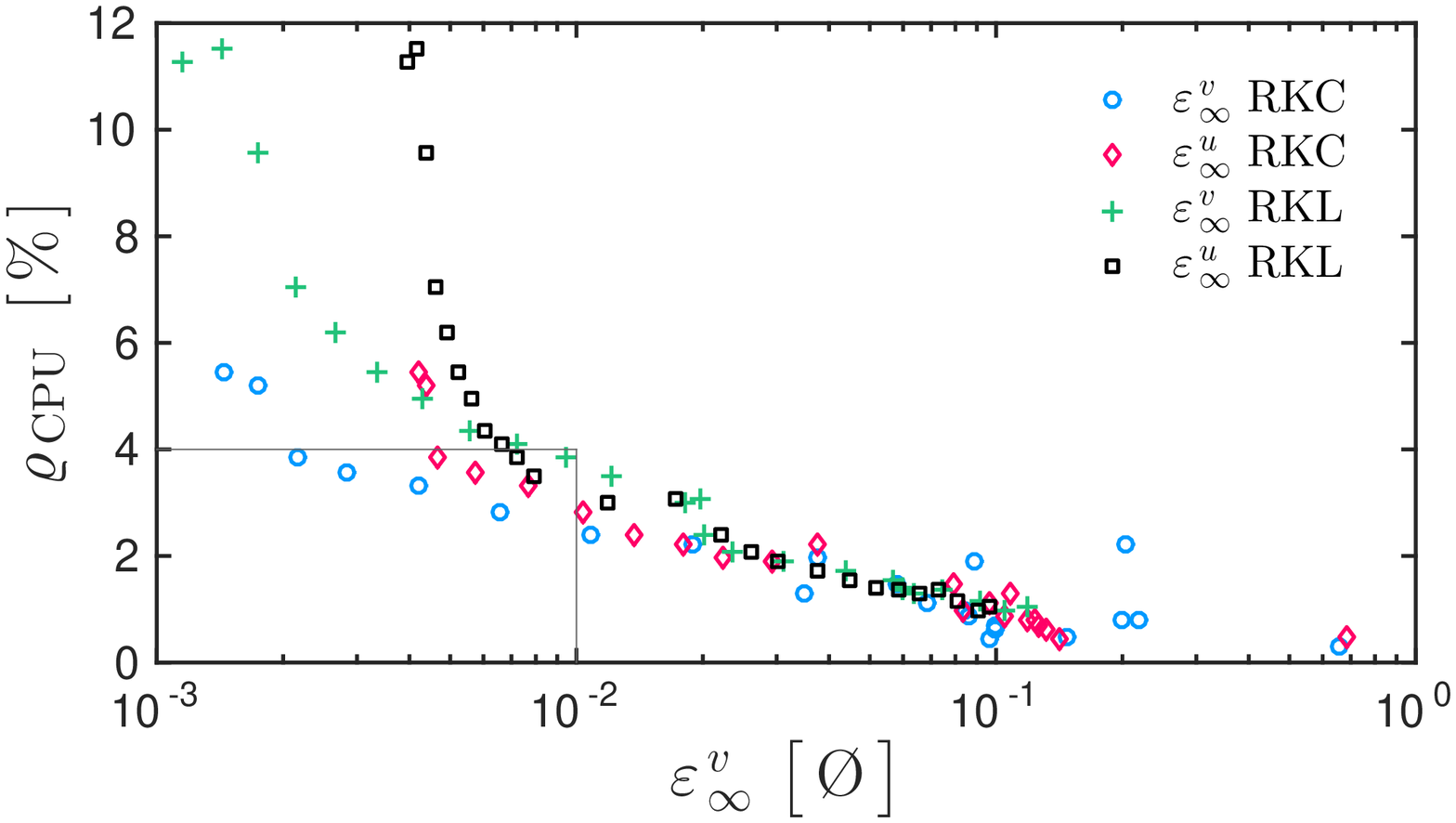}
			\caption{}
			\label{fig:Case_Study_Linear_cpu_ratio}
		\end{subfigure}
		\begin{subfigure}{0.47\textwidth}
			\centering
			\includegraphics[width=1\textwidth]{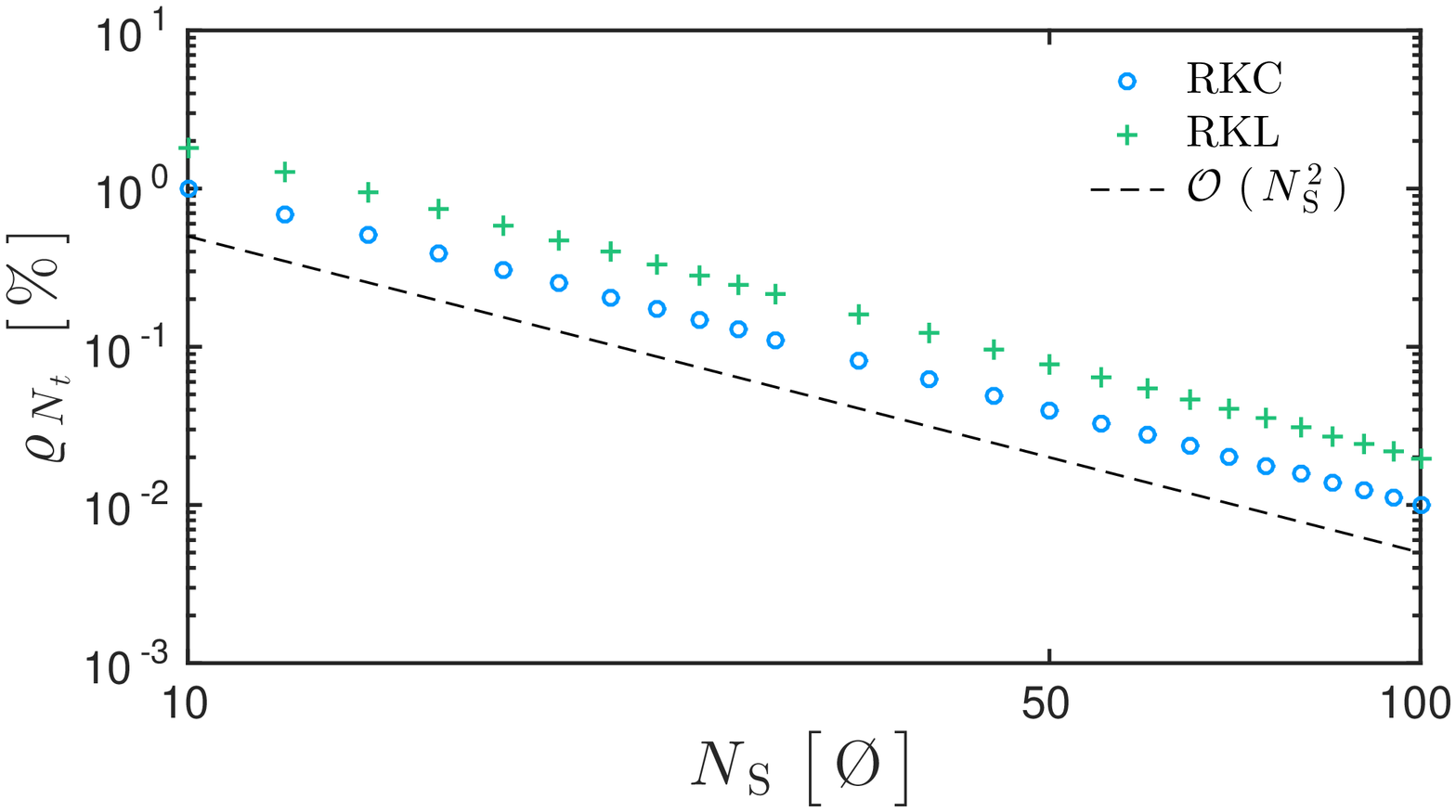}
			\caption{}
			\label{fig:Case_Study_Linear_number_time_step_ratio}
		\end{subfigure}
		\caption{\small Influence of the number of supersteps $N_{\, \mathrm{S}}$ on \rev{$\varepsilon_{\, \infty}$} error for dimensionless $u$ (a) and $v$ (b) variables; on the ratios $\varrho_{\, \text{CPU}}$ (c) and $\varrho_{\, N_{\, \ts}}$ (d) compared to the explicit \textsc{Euler} scheme results. }\label{fig:Linear_case_study_STS_N}
	\end{figure*}

	Results of the simulations provide evidence that the accuracy can be obtained almost at the same level for all schemes. 
	This can be seen from the Figure~\ref{fig:Linear_case_study_error}, where the order of the $\varepsilon_{\, 2}$ error is kept around $\mathcal{O} \, (\,10^{\, -3}\,)$. 
	In addition, from Table~\ref{tab:Linear_case_study} it can be seen that the \rev{$\varepsilon_{\, \infty}$} error is of the same order, being higher for \textsc{Euler} explicit scheme because smaller $\Delta t$ due to requirements of the stability condition.
	Nonetheless, simulations with all methods obtained about two significant correct digits (scd). 
	Important difference between schemes can be noticed from the size of time-steps and the number of time-steps. 
	The CFL stability condition~\eqref{eq:STS_stability} imposes $\Delta t^{\,\star}_{\, \textsc{Euler}}$ to be no bigger than $3.6\times 10^{\,-5}$, which is a really small quantity at the building physics scale. 
	This extreme restriction is relaxed with the STS methods. 
	Even when $N_{\, \mathrm{S}}$ is taken to be equal to $10$, the size of time-step becomes bigger for $100$ times, thereby reducing the number $N_{\, t}$ also by $100$. 
	It basically means that the number of iterations can be reduced considerably and, thus, it is possible to save the extra computational cost. 
	The ratio $\varrho_{\, \text{\tiny CPU}}$ also shows that with STS methods it takes only $5 \,\%$ and $7 \,\%$ of \textsc{Euler} explicit computational time. 
	Hence, at least, it is possible to cut the costs by $93 \,\%$. 
	\begin{table}
		\centering  
		\small
		\caption{\small Comparison of the numerical results for the linear case study. The number of super-time-steps: $N^{\, \text{RKC}}_{\, \mathrm{S}} \egal 10$ and $N^{\, \text{RKL}}_{\, \mathrm{S}} \egal 20$.}                                                                 
		\label{tab:Linear_case_study}                                                                            
		\begin{tabular}{|c||c|c|c|c|}                                                      
			\cline{2-5}                                                                              
			\multicolumn{1}{c|}{}	& \textsc{Euler} & DF & RKC & RKL \\                                                      
			\hline        	\hline                                                                    
			$\Delta t^{\,\star}$ & $3.6\cdot 10^{-5}$ & $10^{-3}$ & $3.6\cdot 10^{\,-3}$ & $7.5\cdot 10^{-3}$ \\               
			\hline                                                                            
			$N_{\, t}$ & $28 \,001$ & $1\, 001$ & $280$ & $133$ \\                                          
			\hline                                                                            
			$\varrho_{\, N_{\,\Delta \ts}} \ \left[\,\%\,\right] $ & $100$ & $3.57$ & $1$ & $0.47$ \\ 
			\hline   	\hline                                                                         
			$\varepsilon_{\infty}\, (\,v\,) $ & $6\cdot 10^{-5}$ & $3\cdot 10^{-3}$ & $3\cdot 10^{-3}$ & $3\cdot 10^{-3}$ \\  
			\hline                                                                            
			$\varepsilon_{\infty} \, (\,u\,) $ & $4\cdot 10^{-5}$ & $3\cdot 10^{-3}$ & $4\cdot 10^{-3}$& $5\cdot 10^{-3}$ \\ 
			\hline                                                                            
			$\text{scd}  \, (\,v\,)$ & $2.45$ & $2.44$ & $2.49$ & $2.13$ \\                           
			\hline                                                                            
			$\text{scd}  \, (\,u\,)$ & $2.69$ & $2.64$ & $2.30$ & $2.04$ \\                           
			\hline 	\hline                                                                           
			$ t_{\, \text{CPU}} \,\left[\,{\sf s}\,\right]$ & $16.3$ & $0.77$ & $1.14$ & $0.89$ \\     
			\hline                                                                            
			$\varrho_{\, \text{CPU}} \ \left[\,\%\,\right]$ & $100$ & $4.72$ & $6.99$ & $5.46$ \\
			\hline                                                                            
		\end{tabular}                                                             
	\end{table}
	The results in Figure~\ref{fig:Linear_case_study_STS_N} are presented to verify the choice of the number of supersteps $N_{\, \mathrm{S}}$. 
	It shows how $N_{\, \mathrm{S}}$ influences the overall efficiency of the simulations.
	The tests are made for $N_{\, \mathrm{S}} \in [\, 10,\,100]$. 

	The global uniform error for dimensionless $u$ (Figure~\ref{fig:Case_Study_Linear_errorSTS_U}) and $v$ (Figure~\ref{fig:Case_Study_Linear_errorSTS_V}) shows that both STS methods follow the order $\mathcal{O} \, (\,N_{\, \mathrm{S}}^{\, 2}\,)$, which scales with the definition of super-time-steps~\eqref{eq:STS_RKC_delta_tS}--\eqref{eq:STS_RKL_delta_tS}.
	The error increases with $N_{\, \mathrm{S}}$, because a bigger $N_{\, \mathrm{S}}$ implies a wider superstep. 
	Hence, fewer discretization points and less accuracy during the simulation is obtained. 
	The ratio $\varrho_{\, N_{\, \Delta \ts}}$ (see Figure~\ref{fig:Case_Study_Linear_number_time_step_ratio}), on the other hand, is decreasing with $N_{\, \mathrm{S}}$. 
	Therefore, depending on the requirements, a bigger $N_{\, \mathrm{S}}$ can be taken to perform faster simulations, but with higher error. 
	In terms of stability, it can be noticed that RKL method is more stable than RKC method. 
	
	Another interesting point to compare is the Figure~\ref{fig:Case_Study_Linear_cpu_ratio} of the ratio $\varrho_{\, \text{CPU}}$ as a function of \rev{$\varepsilon_{\, \infty}$} error. 
	As it can be seen, higher the accuracy, less time is possible to save. 
	For an error at a level of $\mathcal{O} \, (\,10^{\, -2}\,)$ the computational time can be cut for around $96 - 97 \,\%$. 
	By summing up all observations, it can be concluded that RKL method performs more accurately and stable than RKC method for a wider range of $N_{\, \mathrm{S}}$, hence, it is more favorable in practice.  
	\section*{Qualitative comparison with experimental observations}\label{sec:experimental_validation}
	In this section, the main purpose is to validate the reliability of the mathematical model and to estimate the \emph{fidelity} (\cite{clark2010ancient}) of the numerical model. 
	The drying of the wall during its first year after installation shall be simulated and results shall be validated with experimental data.  
	The latter has been obtained from the observations of a house located in Saint-Antoine-l$'$Abbaye, in Is$\grave{\text{e}}$re, South-Eastern France (\cite{soudani2017}). 
	In this house, several walls were built with the RE material, and for the sake of clarity we take only the South wall data. 
	
	One-dimensional simulations have been executed for the RE wall $\ell_{\, \text{RE}} \, = \, 0.5 \,{\sf m} $ in width. 
	Total simulation time is $\tau \, = \, 365 \,{\sf d} \,$, \emph{i.e.} the first year after installation of the wall (starts from July). 	
	The material properties of the RE material are obtained in the previous works by \cite{soudani2017} and presented in Table~\ref{tab:MatProp}. 
	
	The variations of the temperature and moisture contents are measured at $10 \sf{cm}$ from the inside and outside surfaces of the wall (\cite{soudani2017}). 
	Hence, the boundary conditions are taken as \textsc{Dirichlet} type for the shorter width of a wall as $\ell_{\, \text{RE}}^{\, \text{new}} \egal 0.5 \,{\sf m} \moins 2 \times 0.1 \,{\sf m} \egal  0.3 \,{\sf m}$. 
	The initial conditions are $\theta_{\,i} \, = \, 0.53 \ \bigl[\,\varnothing\,\bigr]$ and $T_{\,i} \, = \, 291.3 \,{\mathsf K }\,$.
	
	The space discretization parameter is $\Delta \, x \egal 3 \,{\sf mm} $ for all schemes. 	
	The number of super-time-steps: $N^{\, \text{RKC}}_{\, \mathrm{S}} \, = \, 10$ and $N^{\, \text{RKL}}_{\, \mathrm{S}} \, = \, 20$.
	
\begin{figure*}
		\centering
		\begin{subfigure}{0.47\textwidth}
				\includegraphics[width=1\textwidth]{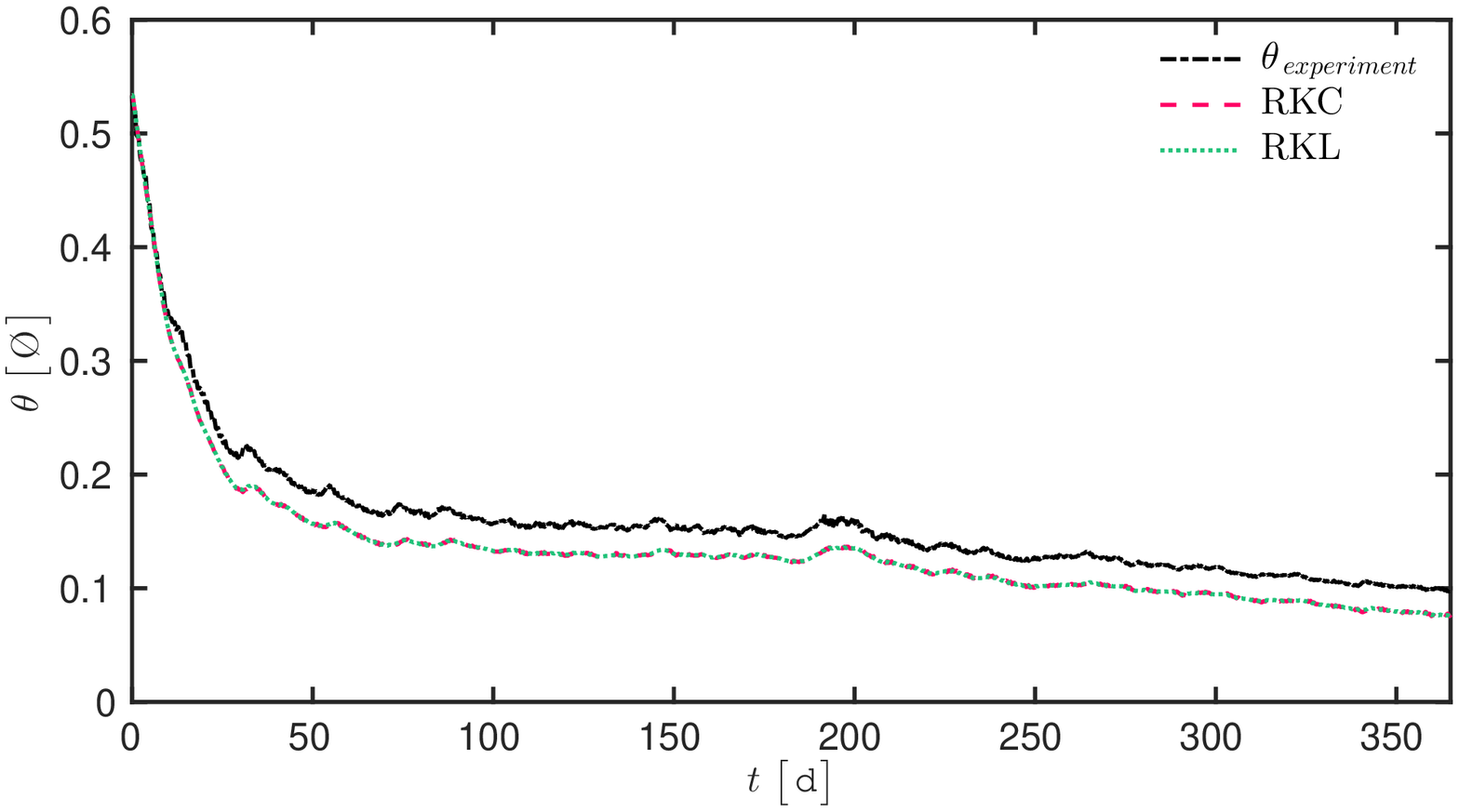} %
		\end{subfigure}
		\begin{subfigure}{0.47\textwidth}
			\includegraphics[width=1\textwidth]{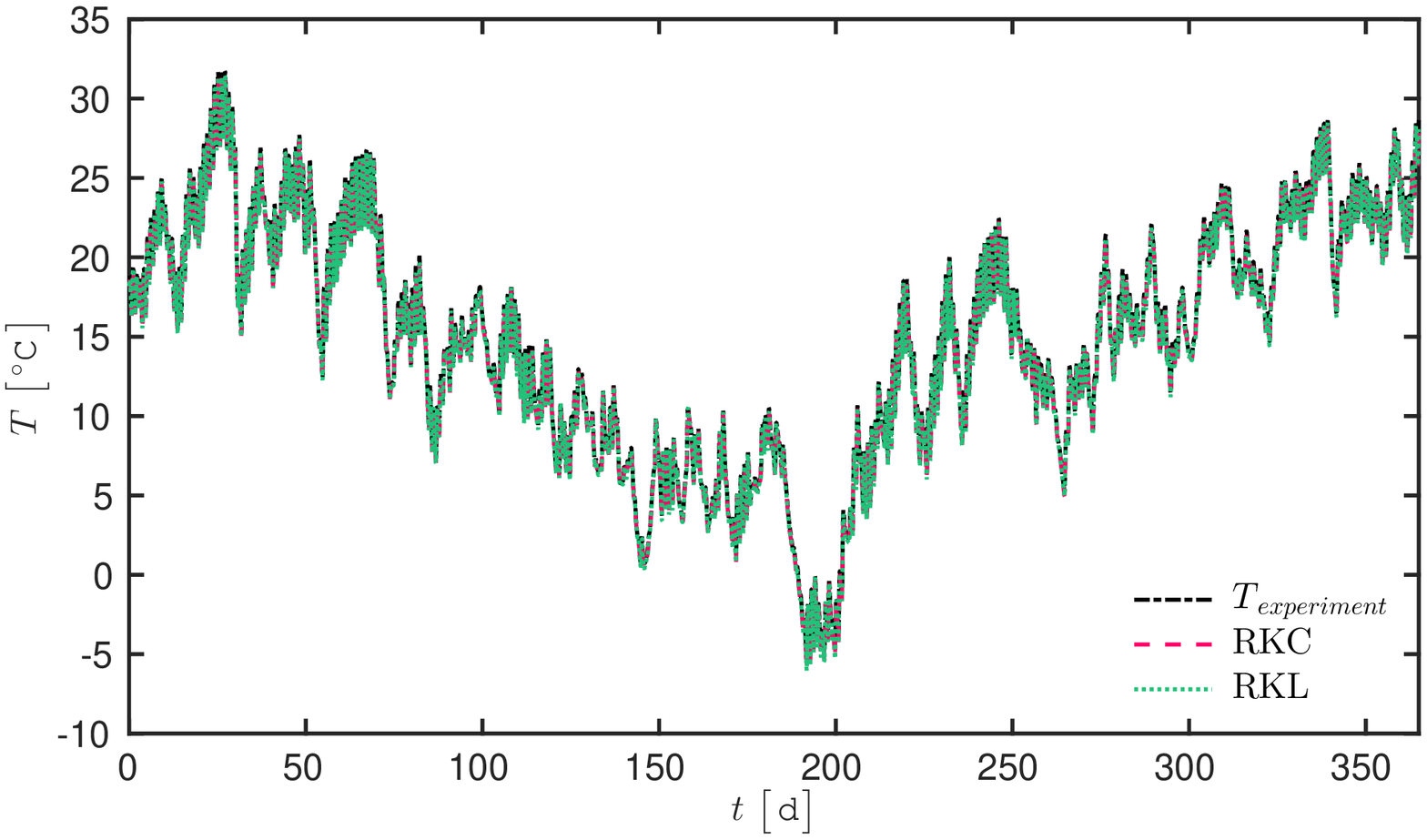} %
		\end{subfigure}
		\caption{\small Time evolutions of the mass content and temperature in the middle of the wall in comparison with the STS schemes along almost one year of the experiment starting from the month of July.}\label{fig:experimental_case_study_Profiles}
	\end{figure*}

	The experimental temperature and the moisture content in the middle of the wall are presented together with the simulation results in Figure~\ref{fig:experimental_case_study_Profiles}.
	As can be seen, the wall considerably dried during first $50$ days (by taking into account that the initial installation was at the end of July). 
	The negative values of the temperature in the winter period are due to the fact that the first year and a half after installation the house was not occupied.
	The general comparison is satisfactory in a qualitative view since some discrepancies can be noted. 
	They arise from a lack of information to model the material properties. 
	Secondly, the rate of drying may depend on variations of the surface transfer coefficients with external factors such as wind, radiation, etc.
	Nonetheless, both mathematical and numerical models proved to be sufficiently reliable to simulate the desired physical phenomena. 
	\section*{Numerical investigation with the physical data}\label{sec:Numerical_Investigation}
	The goal of the numerical study is to analyze the impact of an insulation layer on the moisture state of the RE wall. 
	The properties of a glass wool material have been taken to model the insulation layer (\cite{mendes2008simulation}). 
	The material properties for both layers are displayed in Table~\ref{tab:MatProp}.  
	\begin{table*}
		\centering  
		\small    
		\caption{\small Material properties of the rammed earth and the insulation.}                                            
		\label{tab:MatProp}                                                                  
		\begin{tabular}{|l|c|c|c|c|c|}  
			\cline{2-6}                                                                                
			\multicolumn{1}{c|}{}    &  $D_{\, \theta}$ & $D_{\, T}$ & $\cTs$ & $\kTs$ & $\kTMs$\\                                              
			\hline \hline                                                                       
			RE & $10^{\,-7} \plus 2.4 \cdot 10^{\,-9} \cdot \,(\,\theta \moins 0.1\,) $ & $10^{\, -10}$& $1730\cdot 648 \plus \rho_{\,2} \cdot c_{\,2} \cdot \theta $& $5 \cdot \theta + 0.6$& $4 \cdot 10^{\,-18}$ \\
			\hline \hline                                                                       
			Ins & $10^{\,-20}$  & $0$ & $146 \cdot 840 \plus \rho_{\,2} \cdot c_{\,2} \cdot \theta$ & $0.4875$ & $10^{\,-17}$ \\
			\hline \hline                                                                     
		\end{tabular}                             
	\end{table*} 
    The simulations have been performed for the RE wall $\ell_{\, \text{RE}} \, = \, 0.5 \,{\sf m} $ and the insulation material 
    $\ell_{\, \text{Ins}} \, = \, 0.125 \,{\sf m} $ in width. 	
    The boundary data have been taken as in the previous section as well as the same properties of the RE wall. 
    Total simulation time is $\tau \, = \, 365 \,{\sf d} \,$. 	
	The initial condition for the RE material part is $\theta_{\,i, \, \text{RE}} \, = \, 0.53 \ \bigl[\,\varnothing\,\bigr]$ and for the insulation layer part is $\theta_{\,i, \, \text{Ins}} \, = \, 0.053 \ \bigl[\,\varnothing\,\bigr]$. 
	The initial temperature is $T_{\,i} \, = \, 291.3 \,{\mathsf K }\,$ for both materials.
	
	The space discretization parameter is $\Delta \, x \egal  5\,{\sf mm} $ for all schemes. 
	The number of super-time-steps: $N^{\, \text{RKC}}_{\, \mathrm{S}} \, = \, 10$ and $N^{\, \text{RKL}}_{\, \mathrm{S}} \, = \, 20$.

	One of the interesting points to observe is the drying of the RE material with and without such type of an insulation material. 
	The simulations are performed for three cases: \emph{1)} when an insulation layer is outside (Ins -- RE), \emph{2)} when an insulation layer is inside (RE -- Ins) and \emph{3)} without insulation (RE). 
	
\begin{figure*}
		\centering
		\begin{subfigure}{0.47\textwidth}
			\includegraphics[width=1\textwidth]{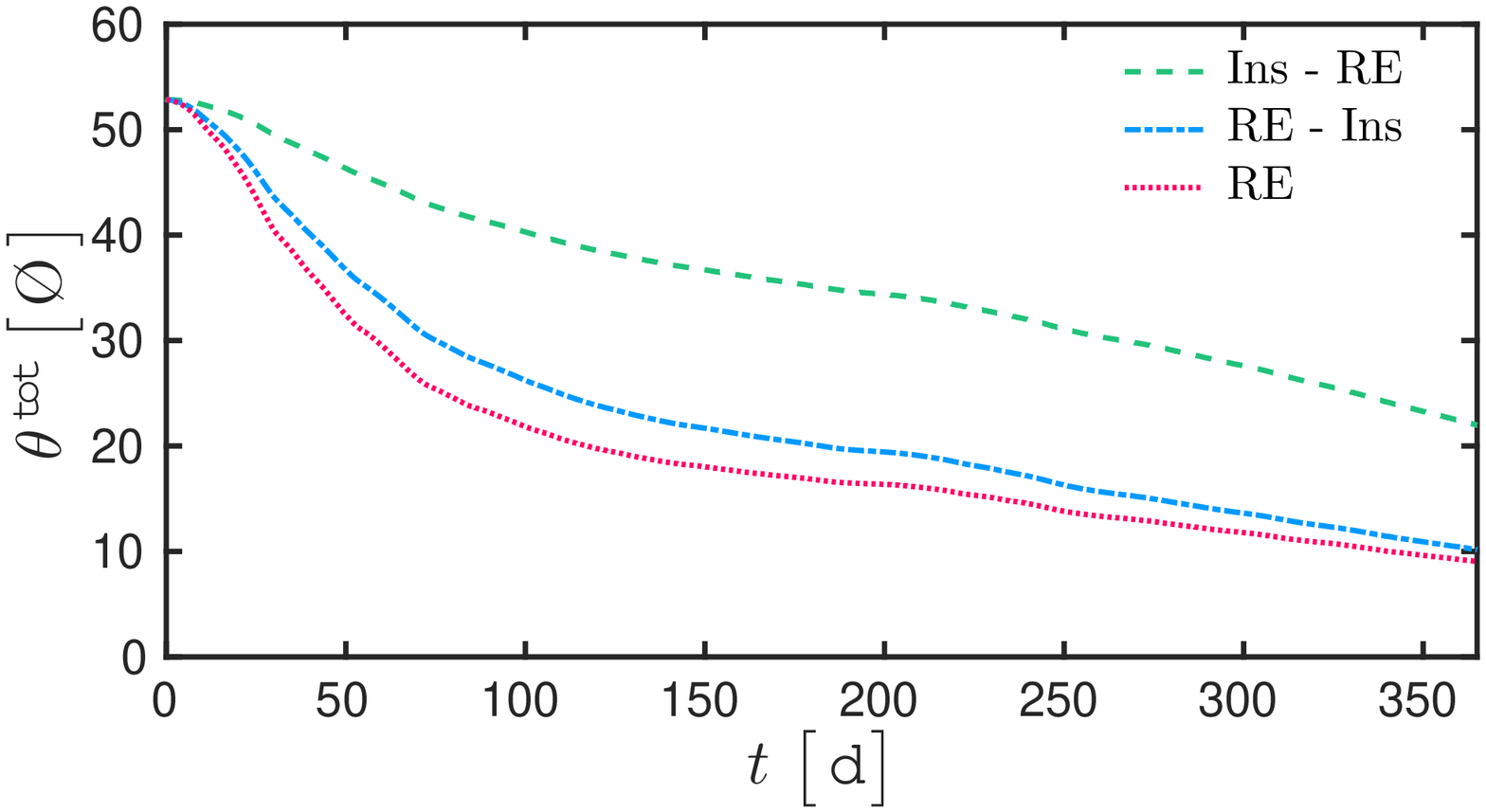} %
		\end{subfigure}
		\begin{subfigure}{0.47\textwidth}
			\includegraphics[width=1\textwidth]{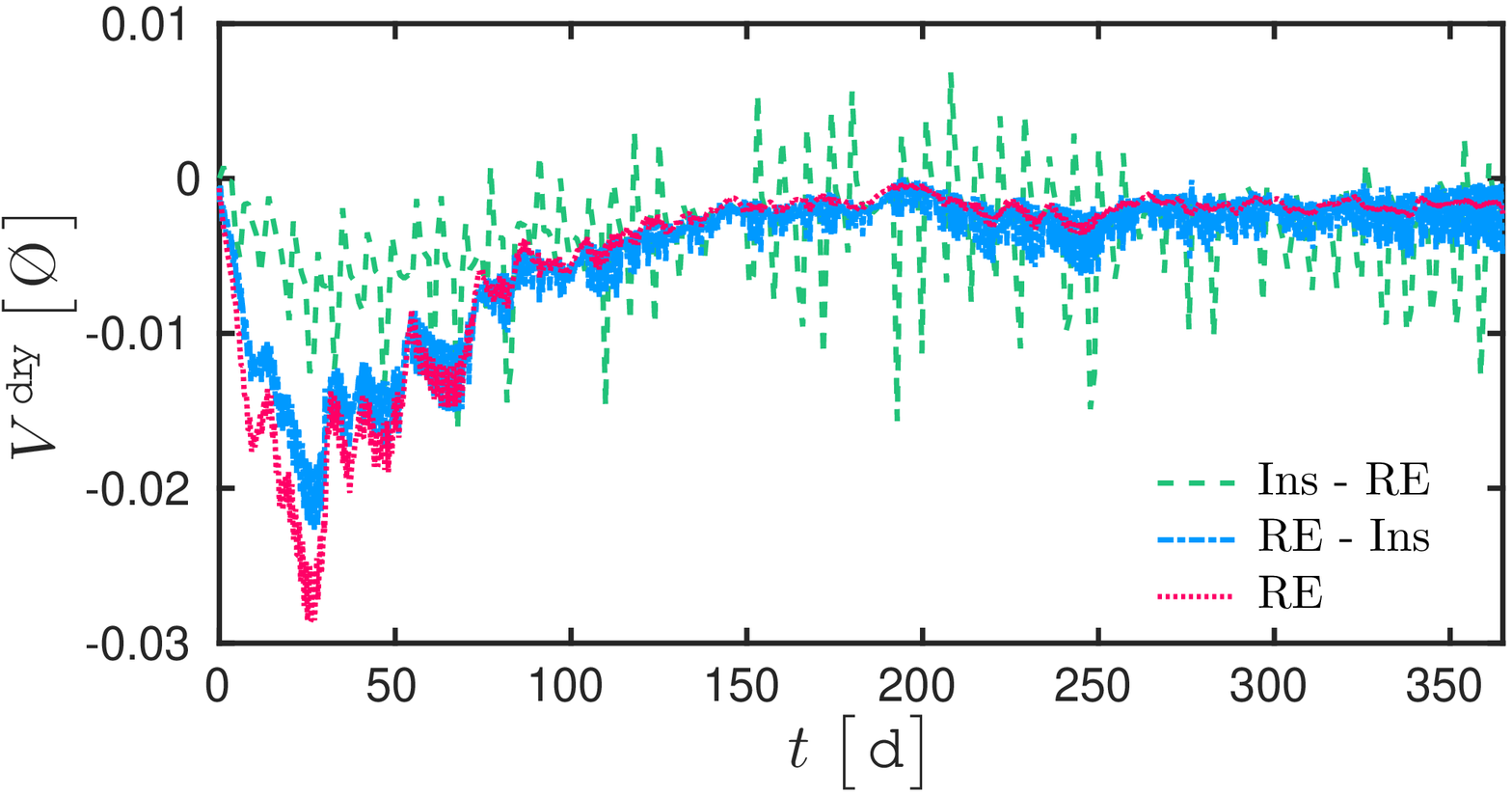} %
		\end{subfigure}
		\caption{\small Total moisture content and the drying  velocity of the rammed earth material. }
		\label{fig:ThetaTOT_Velocity}
	\end{figure*}

	The total moisture content remaining within the material (Equation~\eqref{eq:total_moisture_content}) and the rate of drying (Equation~\eqref{eq:velocity_drying}) for all three cases are plotted in Figure~\ref{fig:ThetaTOT_Velocity}. 
	It can be seen that imposing an insulation layer outside of the wall prevents it from fast drying. 
	This case might be dangerous. 
	In contrast, insulation layer from the inside maintains compatible rate of drying as a wall without insulation. 
	This can also be observed from the rate of drying, where it is alike between cases (RE -- Ins) and (RE). 
	Table~\ref{tab:case_study} is shown to demonstrate the effectiveness of the STS methods in application to a real physical data.
	\begin{table}                                                                             
		\centering 
		\small    
		\caption{\small Numerical results of the STS schemes in comparison with \textsc{Euler} explicit method for one year simulation. The number of super-time-steps: $N^{\, \text{RKC}}_{\, \mathrm{S}} \egal 10$ and $N^{\, \text{RKL}}_{\, \mathrm{S}} \egal 20$.}                                                                 
		 \label{tab:case_study}                                                                               
		\begin{tabular}{|c|c|c|c|}                                                                  
			\cline{2-4}                                                                                   
			\multicolumn{1}{c|}{}& \textsc{Euler} & RKC & RKL \\                                                                          
			\hline  \hline                                                                                   
			$\Delta t \,\left[\,{\sf min}\,\right] $ & $3.4\cdot 10^{\,-2}$ & $3.4$ & $7.1$ \\                        
			\hline                                                                                    
			$N_{\, t}$ & $15 \,629 \, 624$ & $156 \,196$  & $74 \, 379$ \\                                                         
			\hline                                                                                    
			$\varrho_{\, N_{\, \ts}} \ \left[\,\%\,\right] $ & $100$ & $1$ & $0.48$ \\                      
			\hline   \hline                                                                                  
			$ t_{\, \mathrm{\tiny CPU}} \,\left[\,{\sf h}\,\right]$ & $25.4$ & $2.1$& $1.9$ \\                      
			\hline                                                                                    
			$\varrho_{\, \text{\tiny CPU}} \ \left[\,\%\,\right]$ & $100$ & $8.2$ & $7.6$4 \\                      
			\hline                                                                                    
			$\varrho_{\, \text{\tiny CPU}}^{\, \text{\tiny day}} \,\left[\,{\sf s/d}\,\right]$ & $250.6$ & $20.6$ & $19.2$ \\
			\hline                                                                                    
		\end{tabular}                                                    
	\end{table}
	By observing the ratio $\varrho_{\, \text{\tiny CPU}}^{\, \text{\tiny day}}$ it can be clearly seen that STS methods are able to reduce simulation costs by more than $12$ times, and the simulation for a one year period of time might take only two hours instead of a whole one day. 
	In this case again, RKL has slightly better results than RKC, however, the difference in terms of the computational cost is almost negligible comparing to more conventional explicit numerical approaches.    
	\section*{Conclusion}
	The impact of an insulation layer on the behavior of the rammed earth wall has been investigated. 
	The strengths of an innovative STS method are illustrated in comparison with traditional explicit \textsc{Euler} scheme. 
	Results show that the main advantage of the proposed STS method is that it allows to choose at least $100$ times bigger time-steps and to relax considerably stability restrictions. 
	It is also possible to reduce the number of time-steps by more than $200$ times to maintain high accuracy and to cut computational time compared to an explicit scheme for more than $92 \, \%$. 
	In general, it can be concluded that the method proved to be both numerically efficient and accurate enough. 
	Further implementation of such methods to BPS programs is expected to cut computational effort and increase efficiency.	
	
	\section*{Acknowledgment}
	This work was partly funded by the French Environment and Energy Management Agency (ADEME), Technical Center for Buildings (CSTB) and Saint Gobain Isover. 
	The authors also acknowledge the Junior Chair Research program {\it "Building performance assessment, evaluation and enhancement "} from the University of Savoie Mont Blanc in collaboration with the French Atomic and Alternative Energy Center (INES/CEA) and Scientific and Technical Center for Building (CSTB). 
	The authors also would like to thank Dr. \textsc{A. Fabbri}  and Dr. \textsc{L. Soudani} for their valued discussions on the experimental data for the properties of the rammed earth material. 	
	\bibliographystyle{BS2019}
	\bibliography{references}
\end{document}